\documentclass[a4paper]{jpconf}
\usepackage{graphicx}
\usepackage{upgreek}
\usepackage{amsmath}

\begin{document}
\title{GRAAL: Gem Reconstruction And Analysis Library}

\author{R. Farinelli$^{1,2}$,
M. Alexeev$^3$,
A. Amoroso$^3$,
S. Bagnasco$^3$,
R. Baldini Ferroli$^{4,5}$,
I. Balossino$^{1,5}$,
M. Bertani$^4$,
D. Bettoni$^1$,
A. Bortone$^{3,6}$,
F. Bianchi$^{3,6}$,
A. Calcaterra$^4$,
S. Cerioni$^4$,
J. Chai$^3$,
W. Cheng$^3$,
S. Chiozzi$^1$,
G. Cibinetto$^1$,
F. Cossio$^3$,
A. Cotta Ramusino$^1$,
G. Cotto$^{3,6}$,
M. Da Rocha Rolo$^3$,
F. De Mori$^{3,6}$,
M. Destefanis$^{3,6}$,
F. Evangelisti$^1$,
L. Fava$^3$,
G. Felici$^4$,
L. Gaido$^3$,
I. Garzia$^{1,2}$,
M. Gatta$^4$,
G. Giraudo$^3$,
S. Gramigna$^{1,2}$,
M. Greco$^{3,6}$,
L. Lavezzi$^{3,5}$,
S. Lusso$^3$,
H. Li$^3$,
M. Maggiora$^{3,6}$,
R. Malaguti$^1$,
A. Mangoni$^7$,
S. Marcello$^{3,6}$,
M. Melchiorri$^1$,
G. Mezzadri$^1$,
M. Mignone$^3$,
S. Pacetti$^7$,
P. Patteri$^4$,
B. Passalacqua$^{3,6}$,
A. Rivetti$^3$,
M. Savri\'e$^{1,2}$,
S. Sosio$^3$,
S. Spataro$^{3,6}$,
E. Tskhadadze$^8$,
L. Yan$^3$,
R. J. Wheadon$^3$}

\address{
$^1$ INFN Sezione di Ferrara, I-44122, Ferrara, Italy\\
$^2$ University of Ferrara, I-44122, Ferrara, Italy\\
$^3$ INFN Sezione di Torino, I-10125 Turin, Italy\\
$^4$ INFN Laboratori Nazionali di Frascati, I-00044, Frascati, Italy\\
$^5$ Institute of High Energy Physics, Beijing 100049, People's Republic of China\\
$^6$ University of Turin, I-10125 Turin, Italy\\
$^7$ INFN and University of Perugia, I-06100, Perugia, Italy\\
$^8$ Technological Institute of Georgia, Tbilisi, Georgia}

\ead{rfarinelli@fe.infn.it}

\begin{abstract}
MPGD are the new frontier in gas trackers. Among this kind of
devices, the GEM chambers are widely used. The experimental signals acquired with the detector must obviously be reconstructed and analysed. In this
contribution, a new offline software to perform reconstruction,
alignment and analysis on the data collected with APV-25 and TIGER ASICs will
be presented. GRAAL (Gem Reconstruction And Analysis Library) is able to
measure the performance of a MPGD detector with a strip segmented
anode (presently). The code is divided in three parts: reconstruction,
where the hits are digitized and clusterized; tracking, where a
procedure fits the points from the tracking system and uses that
information to align the chamber with rotations and shifts; analysis,
where the performance is evaluated (e.g. efficiency, spatial
resolution,etc.). The user must set the geometry of the setup and then the
program returns automatically the analysis results, taking care of different
conditions of gas mixture, electric field, magnetic field, geometries,
strip orientation, dead strip, misalignment and many others.
\end{abstract}

\section{Introduction}
GEM detector has been invented by F. Sauli in 1997 \cite{Sauli}: it is a gas detector that amplify the primary ionization charge of particles interacting with it.  A GEM is built up by a thin coppered kapton foil with holes of 50 $\upmu$m on the entire surface homogeneously distributed. Using hundreds of Volts between the two GEM faces it is possible to amplify the number of electrons entering in the holes. An electric field outside the GEM foils shifts the electron avalanche to the anode. The anode is segmented by strips and readout by electronics. Three foils of GEM are used to built a triple-GEM and achieve a gain of about 10$^3$-10$^4$.

Triple-GEM detectors have been tested in testbeams (TBs) at CERN and MAMI facilities. To reconstruct and analyse the collected data a dedicated software has been developed. 
This tool is used to:
\begin{itemize}
\item extract the useful information from data such as electronic channel, charge and arrival time of the signal;
\item merge them with the geometrical description of the detector;
\item measure the interaction position of the particles with the detector.
\end{itemize}
The collected information is used to define the performance of the detector such as efficiency and spatial resolution. Different settings have been used to test the triple-GEM performance: different gas mixtures, various electrical field settings inside the detector, magnetic field scan, several angle configuration with respect to the beam, etc. In order reduce the systematic error GRAAL has been automatized to return the characterization of the detector, analysing all the data with the same procedure.

The setup used in the data-taking is composed by at least two detectors under test, a tracking system and a trigger system. Each particle interacting with the setup generates an {\it event}.

\begin{figure}[ht]
\centering
\includegraphics[width=0.3\textwidth]{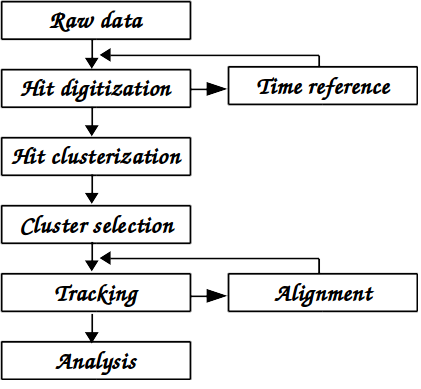}
\caption[Offline software block diagram]{GRAAL block diagram.}
\label{diagramma}
\end{figure}

A schematic diagram of the software is shown in Fig.\ref{diagramma}, where the {\it hit} corresponds to an electronic channel or a single strip; the {\it cluster} to collection of neighbouring strips. The information from the tracking system is used to align the detectors under test. 

\section{Hit digitization}
The performance of a triple-GEM detector depends on the electronics used to readout the signal. In the performed TBs two different kinds of electronics have been used: APV-25 \cite{apv25,apv25_2} and TIGER (Torino Integrated Gem Electronics for Readout) \cite{tiger,tiger_2} chips. APV-25 is a commercial chip that samples the charge up to about 50 fC each 25 ns for 27 times after the trigger signal. TIGER is a 64 channel ASIC with an analogue measurement of charge and time together with a fully digital output. While TIGER can return directly both the charge and time measurements, the signal from the APV-25 needs to be analyzed. The total charge induced on the strip is the maximum charge measured in the 27 time-bins, $Q_{\mathrm{hit}}~=~Q_{\mathrm{max}}$. The time is extracted from the charge behaviour as a function of the time through a Fermi-Dirac (FD) fit on the rising edge of the signal, as described in Eq. \ref{eq:FD_equation}. 
\begin{equation}		
\label{eq:FD_equation}
Q(t)=Q_0+\dfrac{Q_{\mathrm{max}}}{1+\exp \left(- \dfrac{t-t_{\mathrm{FD}}}{\sigma_{\mathrm{FD}}} \right)}
\end{equation}

A pre-analysis is needed as the FD fit requires the initialization of some parameters. $Q_0$ is initialized to the charge mean value of the first three time-bins, $t_{\mathrm{FD}}$ to the mean value between the time associated to $Q_{\mathrm{max}}$ and the time-bin with 10\% of $Q_{\mathrm{max}}$, $\sigma_{\mathrm{FD}}$ to $12.5 \, \mathrm{ns}$. The time related to the signal is assumed to be the inflexion point: $t_{\mathrm{hit}}=t_{\mathrm{FD}}$. A linear fit is used if the FD fit fails or converges to anomalous values. The hit time $t_{\mathrm{hit}}$ is evaluated with the fit in the middle of the rising edge.

\section{Hit clusterization}
The value in the middle of the strip is defined as the strip position, orthogonal to the strip length direction. To measure the incident particle position two different algorithms have been deployed, the charge centroid (CC), that uses the strip charge and position information and the micro-Time Projection Chamber ($\upmu$TPC), that uses also the time information.

The CC method measures the position of the particles with a weighted average as described in the Eq. \ref{eq:CC}.
\begin{equation}		
\label{eq:CC}
x_{\mathrm{CC}}=\dfrac{\sum_{i}^{N_{\mathrm{hit}}}Q_{\mathrm{hit},i} \, x_{\mathrm{hit},i}}{\sum_{i}^{N_{\mathrm{hit}}} Q_{\mathrm{hit},i}}
\end{equation}
where $N_{\mathrm{hit}}$ is the number of hits in the cluster, referred as {\it cluster size}, $x_{\mathrm{hit},i}$ and $Q_{\mathrm{hit},i}$ are the hit position and charge. 

The $\upmu$TPC algorithm reconstructs the particle path by associating to each strip a point in $x_{hits}:z_{hits}$\footnote{z is defined as the beam direction, x and y are orthogonal to the beam} plane, named $\upmu TPC~point$, where $x_{\mathrm{hit}}$ is the strip position and $z_{\mathrm{hit}}$ is the product of the hit time $t_{\mathrm{hit}}$ times the electron drift velocity $v_{\mathrm{drift}}$, computed by Garfield++ \cite{garfield}. A linear fit is performed on the $\upmu$TPC points and the particle position is extracted.

\begin{equation}		
\begin{tabular}{cc}
$z_{\mathrm{hit}}=t_{\mathrm{hit}} \cdot v_{\mathrm{drift}}$~;~$x_{\mathrm{\upmu TPC}}=\dfrac{gap/2 - b}{a}$
\end{tabular}
\label{eq:TPC}
\end{equation}

where $gap$ is the drift gap thickness inside the triple-GEM, $a$ and b the linear fit parameters.

\section{Residual measurement}
The tracking system is used to evaluate the particle path and define the track. From this reconstructed line the expected position $x_{\mathrm{expected}}$ in the test detector plane is evaluated and it is used to measure the residual $\Delta x_{\mathrm{tracker}}$ as shown in Eq. \ref{eq:residual_trk}.

\begin{equation}		
\Delta x_{\mathrm{tracker}} = x_{\mathrm{detector}} - x_{\mathrm{expected}}
\label{eq:residual_trk}
\end{equation}

where $x_{\mathrm{detector}}$ is the position measured by the triple-GEM.
The residual distribution is fitted with a Gaussian function. The $\sigma_{\mathrm{tracking}}$ of the Gaussian would be related to the spatial resolution of the detector but $\Delta x_{\mathrm{tracker}}$ distribution is not used for this purpose because it contains the contributions of the tracking system. $\Delta x_{\mathrm{tracker}}$ is used in the alignment procedures. A different approach has been used to characterize and extract the detector performance, as described in the Sec. \ref{sec:eff}

Using two detectors under test, their relative residual distribution is measured as defined in Eq. \ref{eq:residual_enemy}:

\begin{equation}		
\Delta x_{1,2} = x_{\mathrm{detector},1} - x_{\mathrm{detector},2}
\label{eq:residual_enemy}
\end{equation}

\section{The alignment procedures}
Real detector position and the one reconstructed by the software may have some differences. Alignment procedures are used to remove this mismatch through translations and rotations. Only a sub-sample of the events corresponding to the 10\% of the total event number is used for this purpose. The quantities under study are the shifts along the $x$ and $y$ coordinate, tilts in the $xy$ plane and rotation in the $xz$ plane.

The traslational alignments are measured from the residual distribution $\Delta x_{\mathrm{tracker}}$. The central value of the Gaussian fit is used to shift the detector reference. Then the tracking system itself is aligned using the angular coefficient of the track. Since the tracks must be orthogonal then the entire setup is rotated to match this condition.
 
After this, the alignment related to the detector rotations are needed, starting from the tilt in the $xy$ plane. A bi-dimensional plot $\Delta x_{\mathrm{tracker}}:y_{\mathrm{detector}}$\footnote{$y$ measurement is performed in the same way as the $x$ measurement.} is used. A linear fit of the bi-dimensional distribution is performed and the angular coefficient of the line is used to rotate the detector in the $xy$ plane. Similarly the tilt of the $x$ coordinate of the tested detector with respect to the $x$ coordinate of the first tracker is evaluated. Similarly a bi-dimensional plot is drawn for $\Delta x_{\mathrm{tracker}}:x_{\mathrm{tracker}}$ and a linear fit is used to measure the rotation between the two detectors.

\begin{figure*}[ht!]
\centering
\begin{tabular}{lcr}
\includegraphics[width=0.3\textwidth]{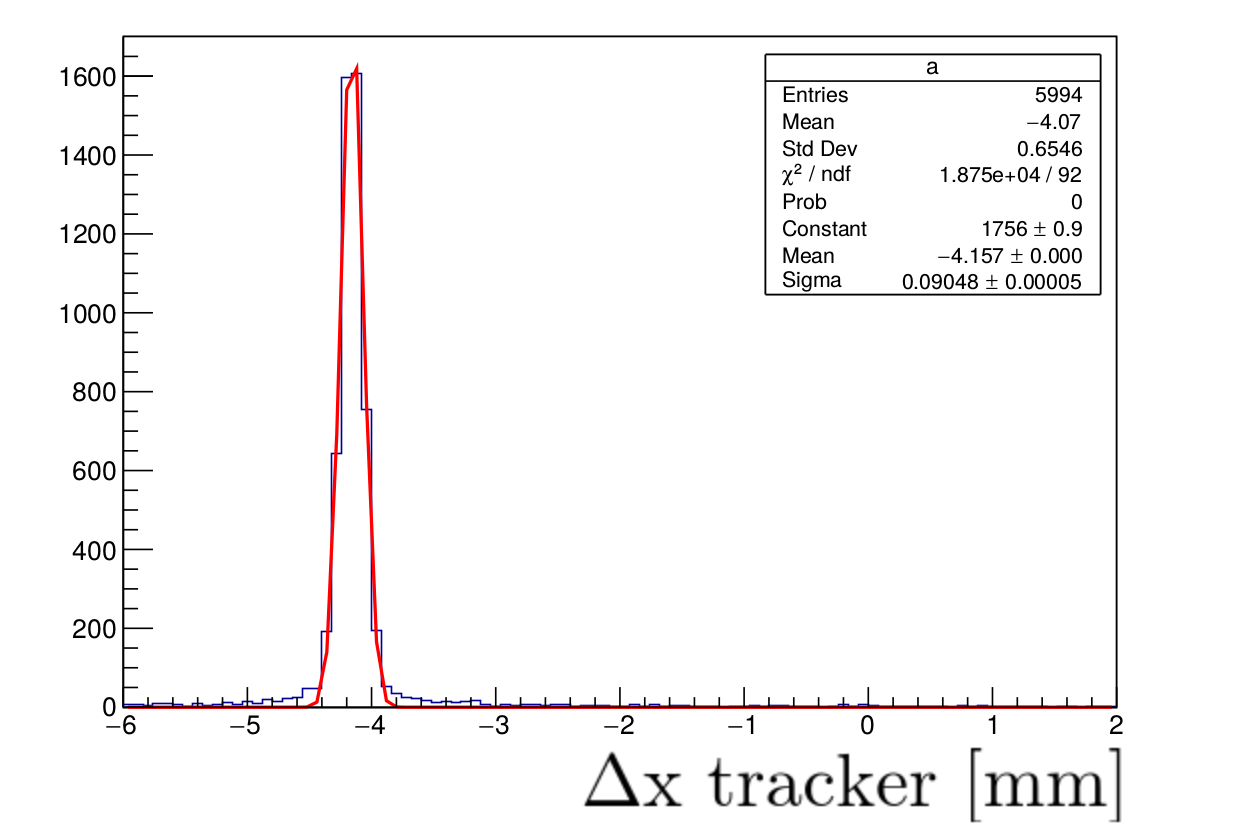} & 
\includegraphics[width=0.3\textwidth]{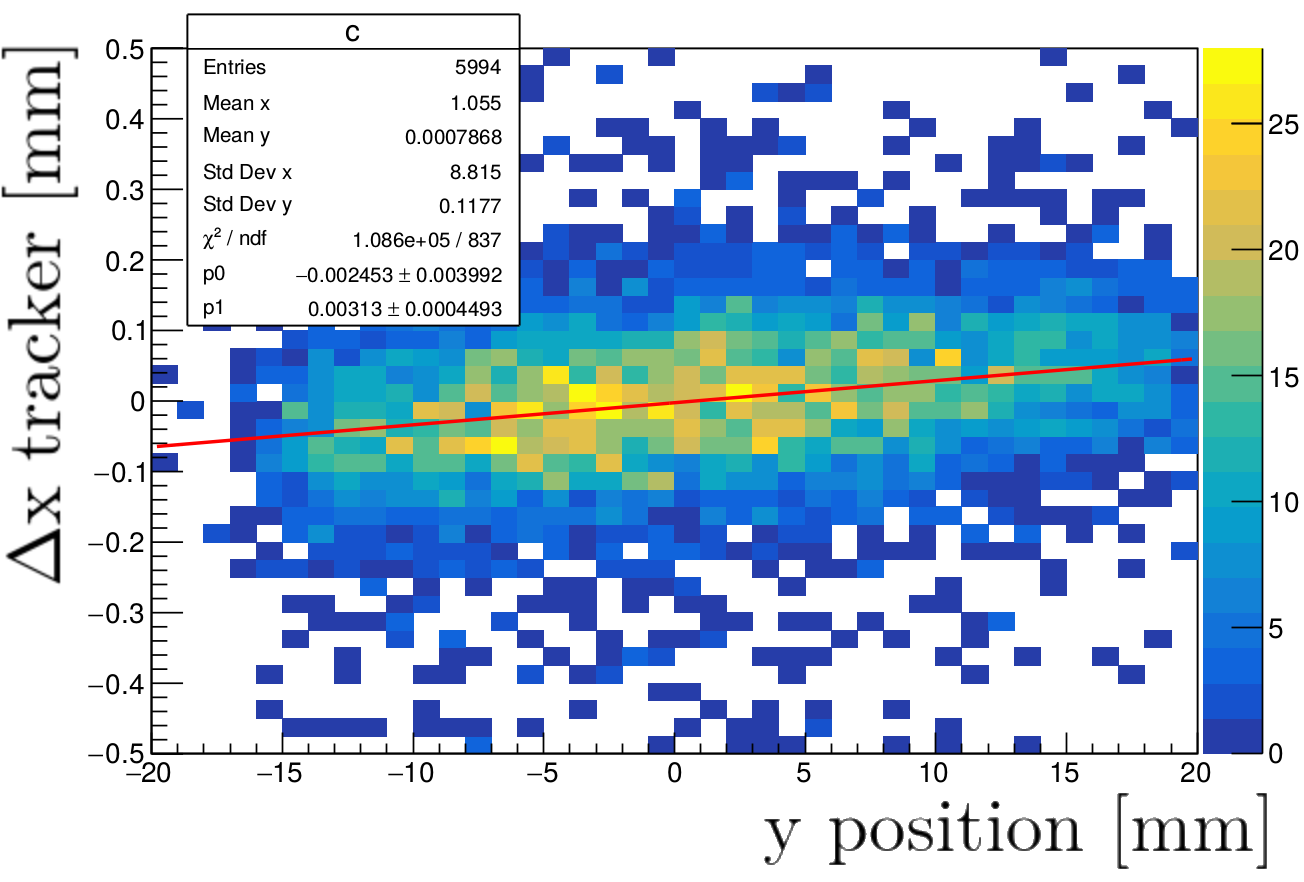} & 
\includegraphics[width=0.3\textwidth]{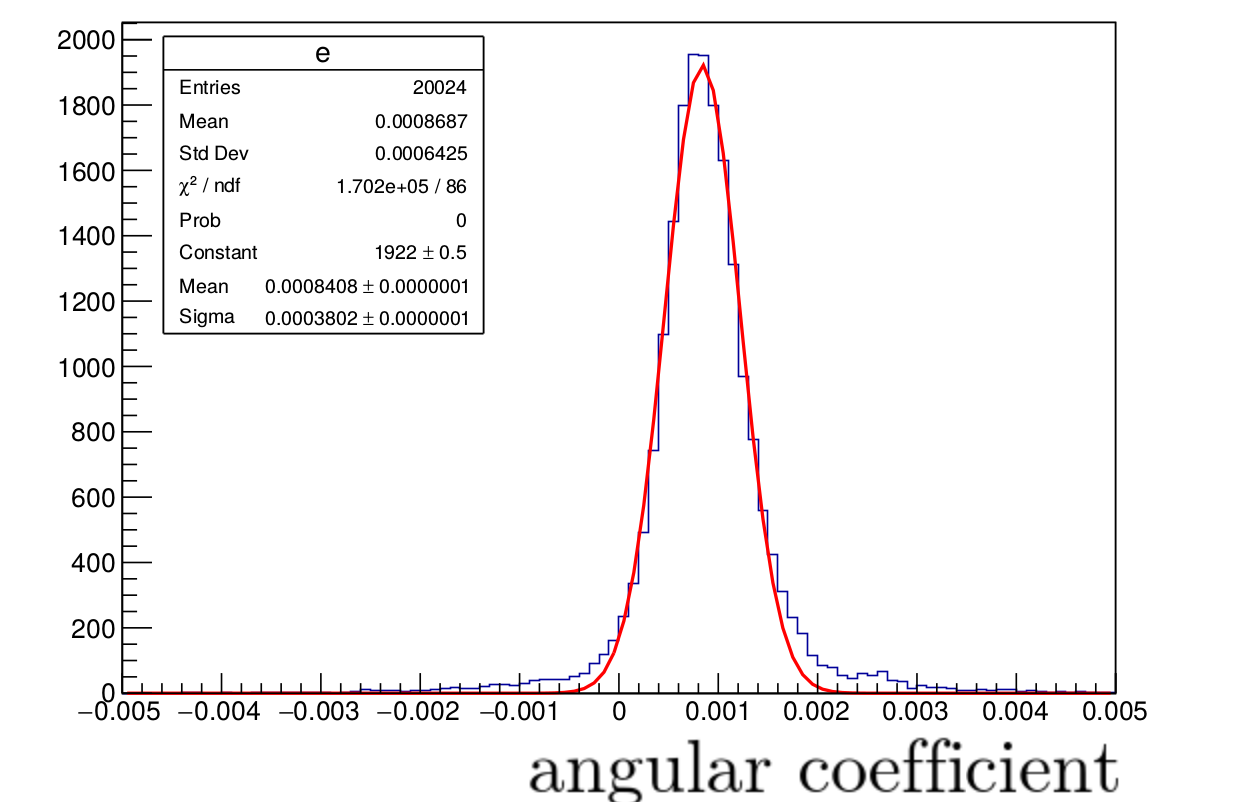} \\
\includegraphics[width=0.3\textwidth]{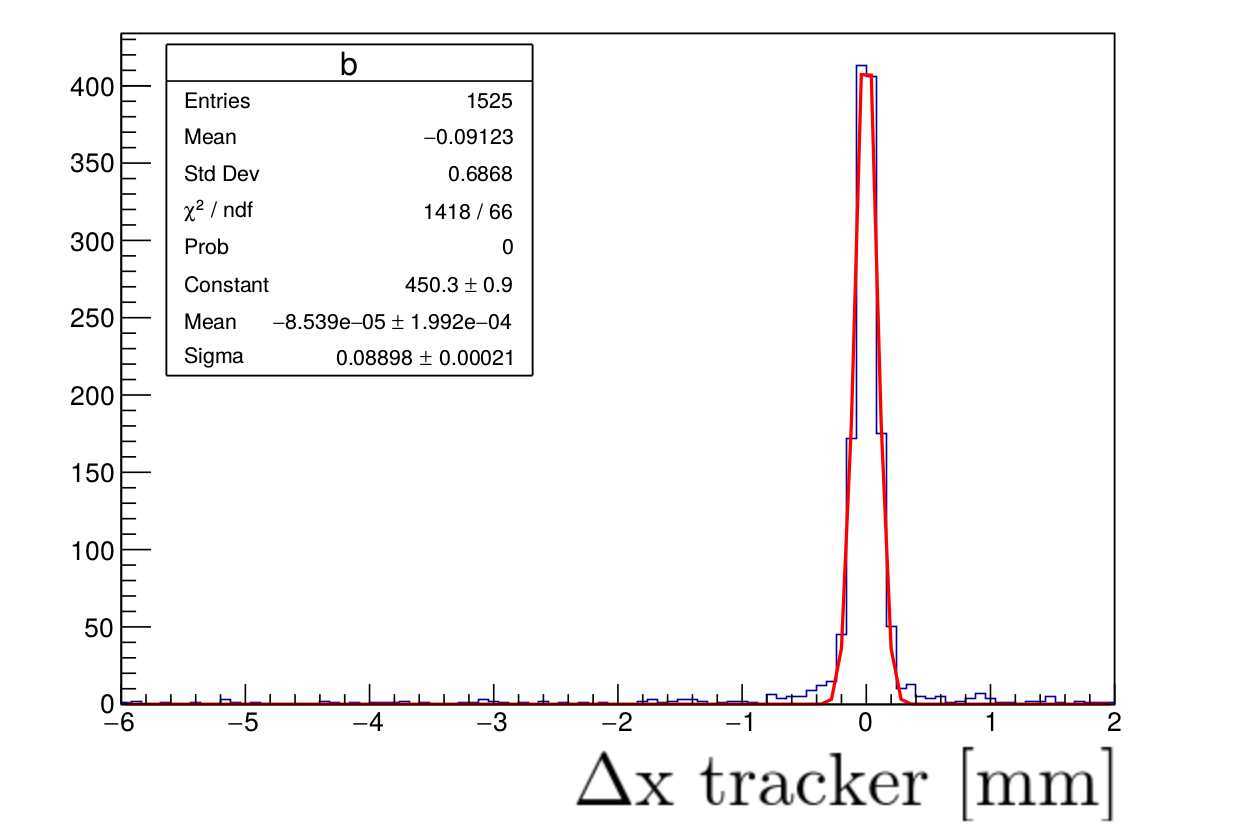} &
\includegraphics[width=0.3\textwidth]{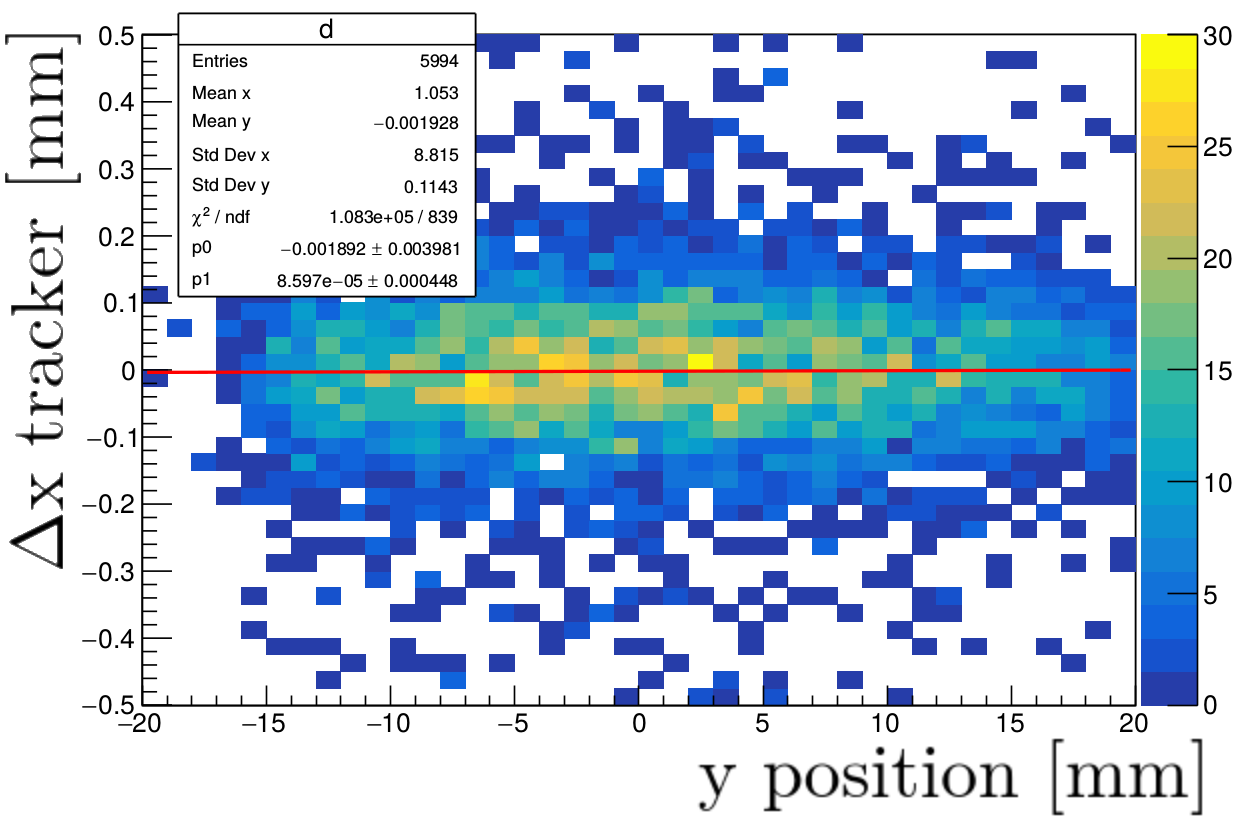} &
\includegraphics[width=0.3\textwidth]{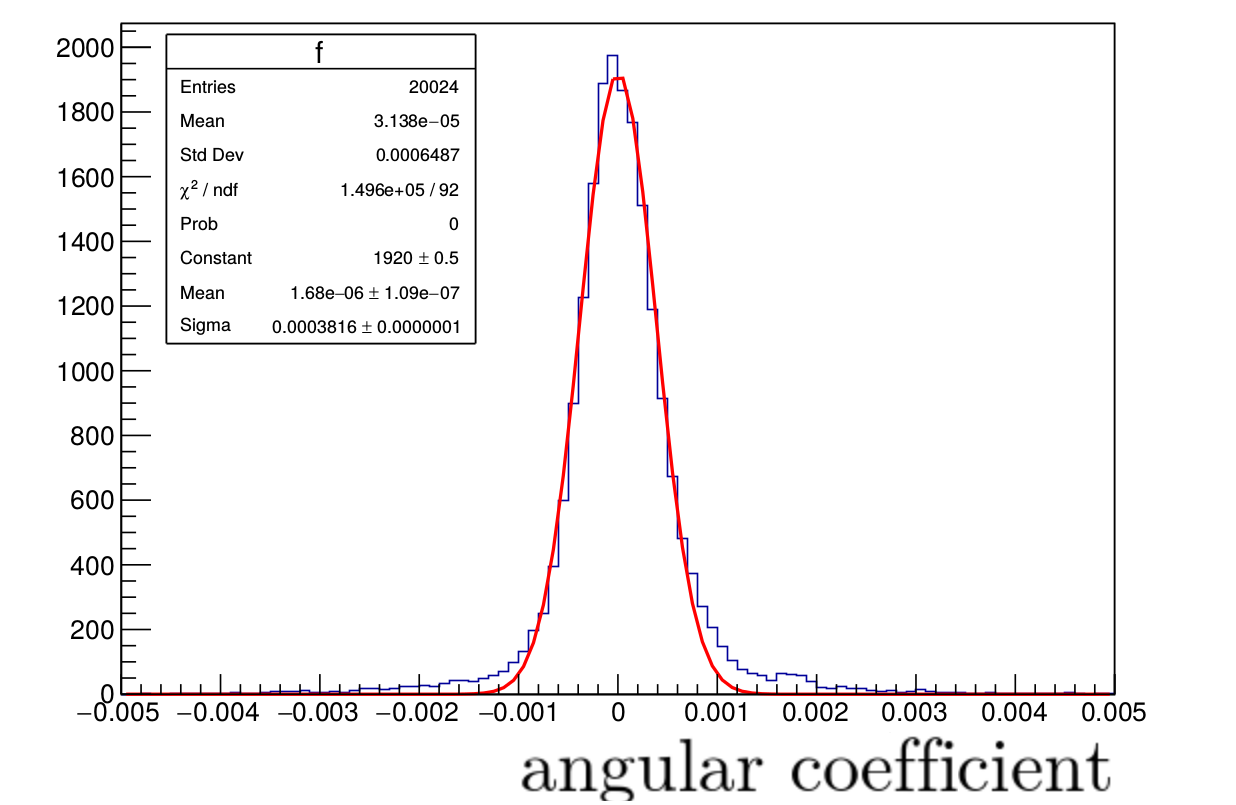}  
\end{tabular}
\caption[Alignment plots]{Alignments plots examples before (top) and after (bottom) the alignment. The first column shows the $\Delta x_{\mathrm{tracker}}$ to remove the shift between the real and reconstructed positions. The second column shows the tilt in $xy$ plane. The third column shows the track angular coefficient measured by the tracking system.}
\label{fig:align}
\end{figure*}

\section{Detector efficiency and signal characterization}
\label{sec:eff}
The performance of the detectors under test is evaluated from $\Delta x_{1,2}$ distribution. A Gaussian fit is used to measure $\sigma_{\mathrm{Gaussian}}$ and the spatial resolution is evaluated as shown in Eq. \ref{eq:resolution_enemy}:

\begin{equation}		
\begin{tabular}{c}
$\sigma_{\mathrm{residual}}^2 = \sigma_{\mathrm{detector},1}^2 + \sigma_{\mathrm{detector},2}^2 $\\$  \sigma_{\mathrm{detector},1} = \sigma_{\mathrm{detector},2}=\sigma_{\mathrm{detector}} \rightarrow \sigma_{\mathrm{detector}} = \dfrac{\sigma_{\mathrm{residual}}}{\sqrt{2}}$
\end{tabular}
\label{eq:resolution_enemy}
\end{equation}

This technique is valid if the two detectors are equal and they have the same performance. This approach removes easily the systematic error due to tracking system on the spatial resolution. To evaluate the efficiency the same distribution $\Delta x_{1,2}$ is used. The number of events within 5$\sigma_{\mathrm{Gaussian}}$ from the mean value is the number of times with a successful reconstruction in the two detectors under test. This number is related to the number of event in which the tracking system has a good event. If the two detectors have the same behaviour then the detector efficiency can be evaluated from Eq. \ref{eq:efficiency}:
\begin{equation}		
\begin{tabular}{c}
$\varepsilon_{1\& 2}=\varepsilon_1 \, \varepsilon_2=\dfrac{N_\varepsilon}{D_\varepsilon}$
,
if $\varepsilon_1=\varepsilon_2=\varepsilon \rightarrow
\varepsilon=\dfrac{N_\varepsilon/D_\varepsilon}{\sqrt{2}}$
\end{tabular}
\label{eq:efficiency}
\end{equation}

where:
\begin{itemize}
\item $D_\varepsilon$ = N$^\circ$ of events with a good reconstructed tracks;
\item $N_\varepsilon$ = N$^\circ$ of event with a good reconstructed tracks and $\Delta x_{1,2}$ within 5$\sigma_{\mathrm{Gaussian}}$ where $\Delta x_{1,2}$ is measured with both detectors under test working properly.
\end{itemize}

\section{Conclusion}
A dedicated software to reconstruct and analyze has been developed to measure the performance of triple-GEM detectors. Charge and time information are used to measure the particle position with CC and $\upmu$TPC algorithms. An alignment procedure is used to remove mismatching between the real detector position and the reconstructed one. The software allows to perform a systematic study of the triple-GEM in different conditions of electrical setting, gas mixtures inside the detector, magnetic field and beam particle rate. The obtained results have been published in several articles \cite{tiger_2,tipp, garzia, emulsioni, rate}. The software has been extended in similar MPGD with an anode segmented by strips \cite{poli}.


\section*{References}
\begin{thebibliography}{9}
\bibitem{Sauli}
F. Sauli, GEM: a new concept for electron amplification in gas detector, \textit{Nucl. Instr. and Meth.} \textbf{A 386}, 531 (1997)
\bibitem{apv25}
L. L. Jones \textit{et al.}, The APV25 Deep Submicron Readout Chip for CMS Detectors, \textit{Conf.Proc.} \textbf{C9909201}, 162-166 (1999) 
\bibitem{apv25_2}
M. J. French \textit{et al.}, Design and results from the APV25, a deep sub-micron CMOS front-end chip for the CMS tracker, \textit{Nucl. Instr. and Meth.} \textbf{A 466}, 359-365 (2001)
\bibitem{tiger}
C. Y. Leng \textit{et al.}, TIGER, a 64 Channel Mixed-mode ASIC for the Readout of the CGEM Detector in the BESIII Experiment, \textit{IEEE NSS conference record} (2016)
\bibitem{tiger_2}
S. Marcello \textit{et al.}, The new CGEM Inner Tracker and the new TIGER ASIC for the BES III Experiment, \textit{PoS EPS-HEP2017} \textbf{505} (2017)
\bibitem{garfield}
R. Veenhof, Garfield, recent developments, \textit{Nucl. Instr. and Meth.} \textbf{A 419}, 726-730 (1998)
\bibitem{tipp}
R. Farinelli \textit{et al.}, A Cylindrical GEM Inner Tracker for the BESIII Experiment At IHEP, \textit{Springer Proc. Phys} \textbf{213} (2018)
\bibitem{garzia}
I. Garzia \textit{et al.}, GEM detector performance with innovative micro-TPC readout in high magnetic field, \textit{EPJ Web Conf.} \textbf{170} 01009 (2018)
\bibitem{emulsioni}
A. Alexandrov \textit{et al.}, High-resolution tracking in a GEM-Emulsion detector, \textit{JINST} \textbf{12} no.09 P09001 (2017)
\bibitem{rate}
L. Lavezzi \textit{et al.}, Performance of the micro-TPC Reconstruction for GEM Detectors at High Rate, \textit{NSSMIC} \textbf{C17}-10-21 (2017)
\bibitem{thesis}
R. Farinelli, Research and development in cylindrical triple-GEM detector with $\upmu$TPC readout for the BESIII experiment, \textit{JINST TH} \textbf{2} (2019) 
\bibitem{poli}
M. Poli Lener \textit{et al.}, The $\upmu$-RWELL: A compact, spark protected, single amplification-stage
MPGD, \textit{Nucl. Instr. and Meth.} \textbf{A 824}, 565-568 (2015)
\end{thebibliography}
\end{document}